# Large nematic susceptibility in the double-Q $C_4$ magnetic phase of Ba$_{1-x}$Na$_x$Fe$_2$As$_2$

Liran Wang,[*] Mingquan He, Frédéric Hardy, Peter Adelmann,
Thomas Wolf, Michael Merz, Peter Schweiss, and Christoph Meingast[†]

*Institut für Festkörperphysik, Karlsruhe Institute of Technology, 76021 Karlsruhe, Germany*

(Dated: 26.3.2018)

The nematic susceptibility of Ba$_{1-x}$Na$_x$Fe$_2$As$_2$ single crystals is studied by measuring the Young's modulus using a three-point-bending setup in a capacitance dilatometer over a wide doping range. Particular emphasis is placed on the behavior within the double-$Q$ antiferromagnetic $C_4$ re-entrant phase. Here, we surprisingly still observe a sizable nematic susceptibility in spite of the well-developed magnetic order, consistent with recent theoretical calculations. Outside the $C_4$ re-entrant region, the behavior is very similar to that of K-doped BaFe$_2$As$_2$. A significant coupling of superconductivity to the shear modulus is observed and is discussed in terms of strong competition between the superconducting and magnetic phases.

The discovery of nematicity in high-temperature superconducting systems has recently attracted considerable attention [1–4]. Superconductivity in iron-based superconductors emerges when the stripe antiferromagnetic spin-density-wave (SDW) transition is suppressed by either doping or pressure [5]. Typically, the SDW is accompanied, or sometimes even slightly preceeded, by a structural distortion from a high-temperature tetragonal ($C_4$) paramagnetic to a low-temperature orthorhombic ($C_2$) symmetry. Here, nematic refers to the structurally distorted phase without long-range magnetic order. It is now fairly clear that this 'nematic' distortion results from an electronically driven instability, involving a complicated entanglement of orbital and spin degrees of freedom [2]. In particular, it was first shown that nematic fluctuations above the SDW transition are of electronic origin by elastoresistivity measurements [6], and signatures of the nematic susceptibility have also been observed in Raman [7, 8], shear-modulus [9–12] and neutron data [13, 14].

Until now, many of the nematic studies have been made on electron-doped 122 systems, however, the hole-doped systems have several interesting features, which deserve attention. In particular, it has now been confirmed that a magnetic $C_4$ re-entrant phase emerges upon doping in all hole-doped 122 compounds studied to date [9, 15–18]. Mössbauer experiments have shown that this phase, in which the orthorhombic distortion is suppressed, results from a superposition of two SDWs with $(0, \pi)$ and $(\pi, 0)$ ordering vectors, and it has therefore been dubbed a double-Q phase. Further, neutron studies have revealed that the magnetic moments point out-of-plane [19], and recent ARPES data provide evidence for a concomitant symmetry allowed charge-order in this phase [20]. In a purely magnetic scenario, these different magnetic phases, including a recently discovered spin-vortex phase [21], are all different manifestations of the same underlying physics [22, 23], although the microscopic reason for the occurence of these phases remains unclear [24, 25] . Clearly, magnetic anisotropy, resulting from significant spin-orbit coupling, is expected to play an important role in these systems and may even introduce additional magnetic phases [26]. One such candidate is the weakly orthorhombic $C'_2$ phase discovered in Ba$_{1-x}$Na$_x$Fe$_2$As$_2$ [18]. Finally, since the magnetic and related nematic fluctuations are likely candidates for the superconducting pairing in these materials, studying the details of these fluctuations is clearly of paramount importance.

In this Letter we study the nematic susceptibility of hole-doped Ba$_{1-x}$Na$_x$Fe$_2$As$_2$ via shear-modulus measurements, with particular emphasis on the re-entrant $C_4$-magnetic phase. The key result of our study is that, unexpectedly, we still observe a sizable nematic susceptibility deep inside the $C_4$ re-entrant phase, in spite of the well-developed magnetic order with spins pointing out-of-plane. This demonstrates that nematic degrees of freedom have not been frozen out in this phase, and our result is consistent with recent theoretical calculations [22], which predict an enhanced nematic susceptibility. These findings may shed light on the general relation between superconductivity and nematic fluctuations and/or nematic quantum criticality [27–29]. Finally, the high-resolution of our technique allows a detailed study of the coupling of superconductivity to the nematic fluctuations, which is discussed in terms of a strong competition between superconducting and magnetic phases.

Single crystals of Ba$_{1-x}$Na$_x$Fe$_2$As$_2$ were grown from flux and characterized using 4-circle single-crystal x-ray diffraction, as described in [18]. Young's modulus measurements were carried out in a capacitance dilatometer using a three-point-bending technique [9, 30]. A schematic sketch of this setup, in which the force from the dilatometer springs causes a deflection of the crystal, is shown in Fig. 1b. The crystals were cut and oriented such that the $[110]_{tet}$ direction of the high-temperature tetragonal cell is perpendicular to the beams of the three-point fixture (see inset of Fig. 1b). In this orientation the bending measurement couples to the soft mode of the magnetic (nematic) phase transition [30]. Details of the measurements are similar to those reported in Ref. [9].

---
[*] liran.wang@kit.edu
[†] christoph.meingast@kit.edu

Supplemental thermal-expansion measurements on the same crystals were carried out using the same dilatometer [18].

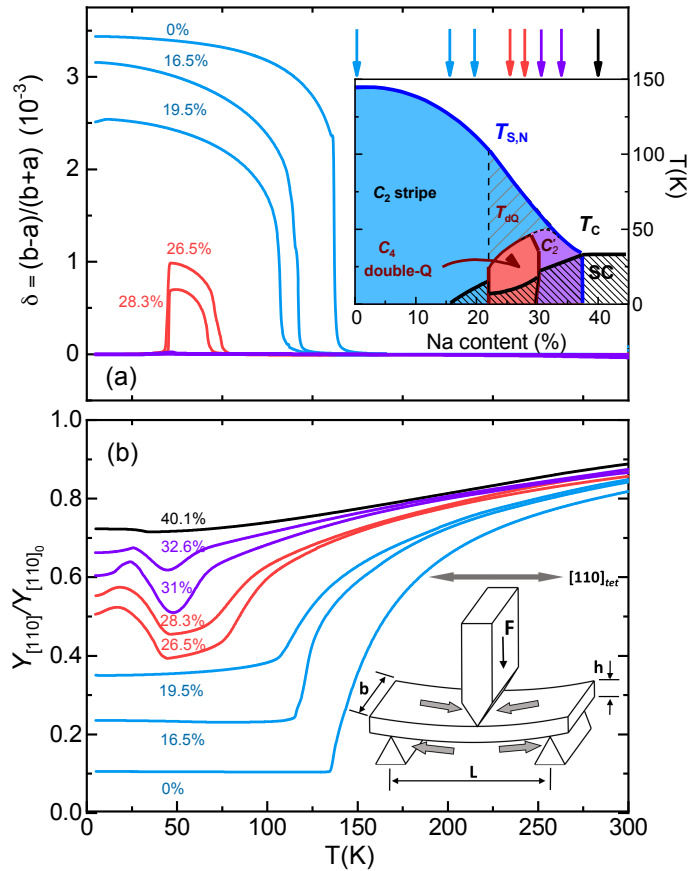

FIG. 1. (a) Orthorhombic distortion versus temperature of $Ba_{1-x}Na_xFe_2As_2$ single crystals determined using capacitance dilatometry. The inset shows the phase diagram from Ref. [18] and the Na concentrations for the present measurements (arrows on top of inset). (b) Corresponding normalized Young's modulus, $Y_{[110]}/Y_{[110]_0}$, from static three-point bending measurements on the same crystals used for dilatometry.

The inset of Fig. 1a shows a schematic phase diagram of $Ba_{1-x}Na_xFe_2As_2$ [18], in which the Na contents of the single crystals used in this paper are indicated by the arrows. In Figures 1a and 1b we present the orthorhombic distortion $\delta \equiv (b-a)/(b+a)$ and the normalized Young's modulus $Y_{[110]}/Y_{[110]_0}$ for these crystals, respectively. In order to concentrate on the electronic effects upon the elastic properties, we have subtracted a non-critical background due to phonon anharmonicity and have normalized the Young's modulus data by the unperturbed modulus $Y_{[110]_0}$, as detailed in Ref. [9]. As shown previously [9, 30], the Young's modulus in these systems represents a good approximation of the shear modulus, $C_{66}$, determined via ultrasound measurements [12]. The general response of $Y_{[110]}/Y_{[110]_0}$ above $T_{s,N}$ upon temperature and doping is very similar to previous data on K-doped Ba122 [9], except in the region of the phase diagram where the $C_4$ and $C_2'$ phases occur (26.5 - 32.6% Na). In particular, we find the same Curie-Weiss softening of the shear modulus for low-doping levels (0 % - 20 %) and a less critical softening for higher Na contents (26.5 - 32.6 %). Further, just as in optimally or overdoped K-and Co-doped systems, a clear hardening below $T_C$ is observed in the crystal without magnetic/structural order (40.1 % Na - black curve). The occurrence of the $C_4$ re-entrant phase is clearly observed by the sudden disappearance of the orthorhombic distortion, $\delta$, around 50 K (red curves in Fig. 1a) and results in a hardening of $Y_{[110]}/Y_{[110]_0}$. Although the transition to the $C_2'$ phase (purple curves) is not visible (on this scale - see [18] for details) in the orthorhombic distortion (Fig.1a), very clear signatures of the transition are observed in $Y_{[110]}/Y_{[110]_0}$ (Fig. 1b). Finally, we also note that nematic and magnetic transitions occur simultaneously in both K- or Na- doped systems at $T_{s,N}$ [15, 18, 31].

As argued previously [9], the static three-point-bending method does not provide physically sensible results below $T_{s,N}$ in the $C_2$ magnetic phase. On quite general grounds, one expects a hardening of the shear mode below $T_{s,N}$ instead of the constant values shown in Fig. 1b [30], and dynamic three-point-bending measurements [30], inelastic neutron scattering [13], as well as resonant ultrasound data (Carpenter et al, unpublished) indeed show this expected hardening below $T_{s,N}$. The low constant value in our measurements is due to the formation of twins with a sizable orthorhombic distortion which are aligned by the applied force such that the crystal is spontaneously 'bent', which is analogous to the finite magnetization of a ferromagnet below the Curie temperature in an applied field[32].

In the following we argue that our three-point bending data do in fact provide sensible results within the $C_4$ region, in spite of the above limitations. This is because the tetragonal symmetry is restored in this phase [15, 18], and thus the aforementioned problems of twins in the $C_2$ phase no longer apply. Further, we demonstrate the extreme sensitivity of the system in this region of the phase diagram to small stresses applied by the dilatometer.

Fig. 2a shows how the elastic response of the 26.5 % crystal changes with magnitude of the applied force. Here, the force was tuned by changing the starting gap of the capacitance dilatometer, and the applied force is quantified in terms of the maximum tensile (compressive) stress in the sample as follows:

$$\sigma_{max} = 3FL/2bh^2$$

where $F$ is the applied force, $L$ is the length between the support beams, $b$ is the width, and $h$ is the thickness of the sample (see inset of Fig. 1b). We find that the elastic response is independent of the applied force in both $C_4$ phases, while it changes drastically within the $C_2$ phase. This demonstrates the reliability of our method in the tetragonal phases. Figures 2a and 2b also show that the elastic response at $T_{dQ}$ and $T_{s,N}$ is quite broad, in is expected to have the same effect, since it also favors the orthorhombic $C_2$ phase, however with the opposite domain orientation. In the three-point bending experiment, the crystal experiences both compressive, as well as tensile stresses, as indicated in the inset of Fig. 1b. Additionally, the stress in the bending setup is very inhomogeneous, changing from compressive to tensile in the middle of the sample, and is thus expected to lead to a significant broadening, as well as the shifting of $T_{dQ}$ to higher and $T_{s,N}$ to lower temperatures, as experimentally observed (see Fig. 2).

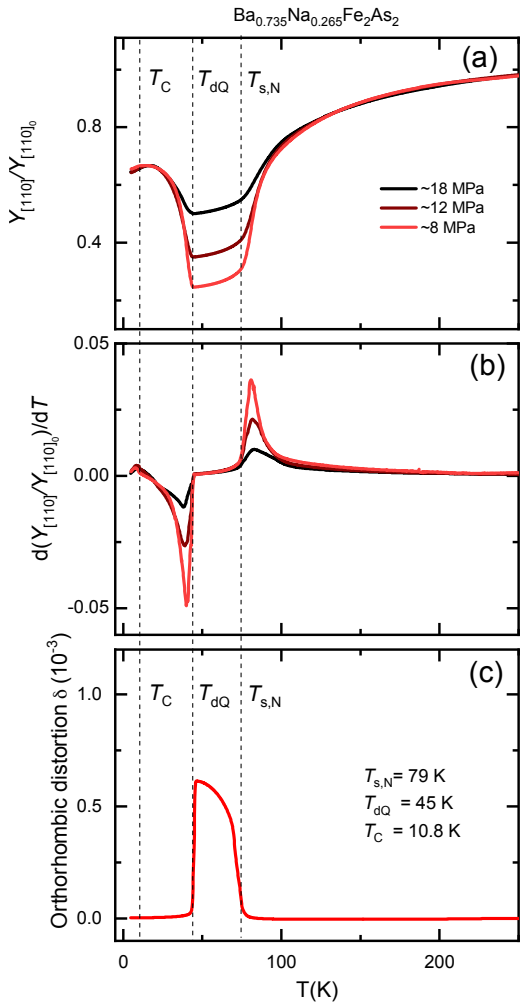

FIG. 2. (a) Normalized Young's modulus, $Y_{[110]}/Y_{[110,0]}$, of $Ba_{0.735}Na_{0.265}Fe_2As_2$ under different maximal compressive (tensile) stress conditions. (b) The temperature derivative of $Y_{[110]}/Y_{[110,0]}$. (c) The corresponding orthorhombic distortion, $\delta \equiv (b-a)/(b+a)$, of the same crystal. The dashed lines indicated the transition temperature $T_{s,N}$, $T_1$ and $T_C$. (see text for details)

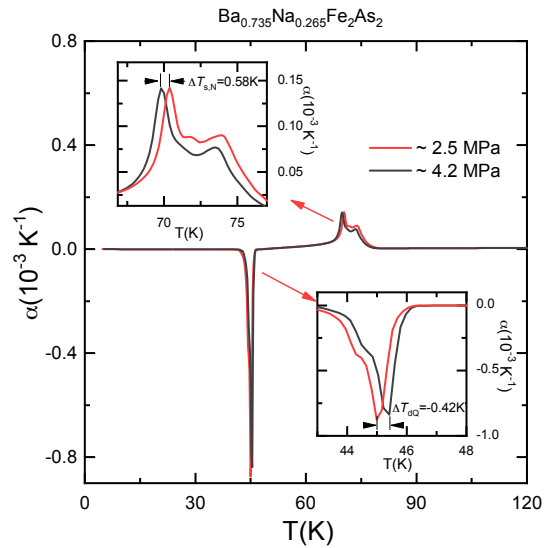

FIG. 3. Thermal expansion coefficient, $\alpha(T)$, along $[110]_{Tet}$ direction of $Ba_{0.735}Na_{0.265}Fe_2As_2$ under two different uniaxial pressures. The inserts illustrate the pressure induced shift of the transition temperatures $T_{s,N}$ and $T_1$, respectively.

contrast to the quite sharp transitions observed in the orthorhombic distortion (Fig. 2c). This broadening results from the inhomogenous strain in the three-point-bending setup and the very large pressure dependences of these transition temperatures, as explained in the following.

Both magnetic transitions are extremely sensitive to uniaxial stress as demonstrated in Fig. 3, in which the thermal expansion along the $[110]_{Tet}$ direction for two different uniaxial pressures applied by the dilatometer is shown. We find $dT_{s,N}/dp = +\,340$ K/GPa and $dT_1/dp = -\,250$ K/GPa, both of which are extremely large values and of opposite sign. As expected, a uniaxial compressive stress stabilizes the orthorhombic $C_2$ phase by increasing $T_{s,N}$ and decreasing $T_{dQ}$. We note that a tensile stress

In Fig. 4 we present our key result, i.e. the nematic susceptibility, $\chi_\varphi(T)$, derived from our Young's modulus data using a mean-field Landau expansion of the free energy, as described in detail in [9, 30]. Note that the the nematic susceptibility is expressed as $\lambda^2 \chi_\varphi / C_{66,0}$, where $\lambda$ is the electron-lattice coupling constant and $C_{66,0}$ the bare elastic constant [9]. Here, the data within the $C_2$ phase region are not plotted, for the reasons discussed above. The dashed lines serve as a guide to the eye for the interpolated behavior, assuming that there would be no $C_2$-magnetic phase. At the low doping range, samples with 0%, 16.5% and 19.5% Na, exhibit typical Curie-Weiss behavior (black dashed lines), as previously observed for K-doping [33].

The behavior changes drastically for x = 0.26, where the $C_4$ re-entrant magnetic phase sets in. Here, the nematic susceptibility no longer diverges to low temperature and resembles the behavior at optimal doping, although with a significantly larger low-temperature value of $\chi_\varphi$. The important result here is the relatively large values of the nematic susceptibility within both the $C_4$

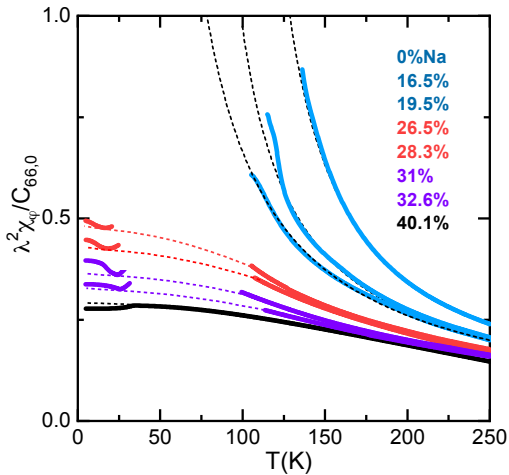

FIG. 4. Nematic susceptibility, $\chi_\varphi$, in units of $\lambda^2\chi_\varphi/C_{66,0}$. The black dotted lines are Curie-Weiss fits in the underdoped region. The dashed colored lines are guide to the eye.

and $C_2'$ regions of the phase diagram. We note that a value of $\lambda^2\chi_\varphi/C_{66,0} = 0.5$, as observed for the 26.5 % Na crystal, represents one half of the critical value expected at the nematic transition [9]. These large values clearly demonstrate that the nematic degrees of freedom are still operating within the re-entrant magnetic phases, as has also been predicted by theory [22]. In a sense, this is a confirmation of the spin-nematic scenario, in which the different magnetic phases are just different facets of the same underlying physics [22, 23, 26, 34]

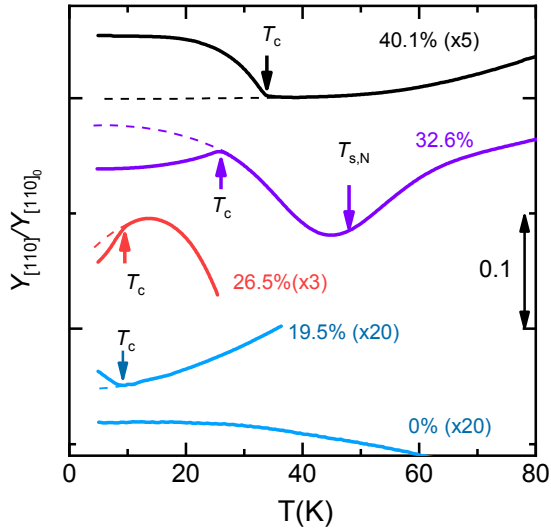

FIG. 5. Magnified view of the shear modulus response near $T_C$ for various Na concentrations.

Finally, we present in Fig. 5 a magnified view of the shear modulus response near $T_C$ for various Na concentrations. The dash lines present the estimated signal without the superconducting transition. A clear hardening below $T_C$ is observed for crystals with 19.5 % Na and 40.1 % Na, similar to the data in K-doped Ba122 [9]. In contrast, within the $C_4$ and $C_2'$ region of the phase diagram, a significant softening occurs below $T_C$. Previously, a large suppression of superconductivity within the $C_4$ re-entrant phase has been reported both for K-Ba122 and Na-Ba122 [9, 18], indicating a strong competition between this magnetic phase and superconductivity. In particular, the $C_4$- re-entrant phase in K-Ba122 actually reverts back to the stripe $C_2$ phase below the superconducting transition [31], which is a clear indication for the preference of superconductivity for the stripe phase over the $C_4$ re-entrant phase. Similarly, the softening below $T_C$ observed in the present shear-modulus data is an indication for a preference of superconductivity for stripe-type magnetism, since a softening below $T_C$ corresponds to an increase in the orthorhombic distortion in our three-point-bending setup at finite pressure. Alternatively, it has been proposed that a softening of the shear modulus may be considered as an indication of a possible proximity to a transition between $s$- and $d$-wave states [27]. We note that an unusual enhancement of the nematicity below $T_C$ has actually been observed in $Fe(Se_{1-x}S_x)$ [35].

In summary, we have studied the nematic susceptibility across the Na-doped Ba122 phase diagram. Outside the double-Q phase region, the behavior is found to be very similar to that of K-doped $BaFe_2As_2$ [9, 30]. Deep inside the $C_4$ re-entrant magnetic phase we still observe a sizable nematic susceptibility, in spite of the well-developed magnetic order. This demonstrates that nematic degrees of freedom have not been frozen out in this phase, and our result is consistent with recent theoretical calculations [22], which predict an enhanced nematic susceptibility. Our finding is expected to be relevant in particular with respect to the relation between superconductivity and nematic fluctuations (e.g. nematic quantum critical point) [27–29]. Finally, we observe a softening of the shear mode below $T_C$ within the $C_4$ re-entrant phase, which is attributed to the strong competition between superconductivity and this magnetic phase[18].

We acknowledge enlightening discussions with Qimiao Si.

L. W. and M. H. contributed equally to this work.

[1] E. Fradkin, S. A. Kivelson, M. J. Lawler, J. P. Eisenstein, and A. P. Mackenzie, Annual Review of Condensed Matter Physics **1**, 153 (2010).
[2] R. M. Fernandes, A. V. Chubukov, and J. Schmalian, Nat Phys **10**, 97 (2014).
[3] A. J. Achkar, M. Zwiebler, C. McMahon, F. He, R. Sutarto, I. Djianto, Z. Hao, M. J. P. Gingras, M. Hücker, G. D. Gu, A. Revcolevschi, H. Zhang, Y.-J. Kim, J. Geck, and D. G. Hawthorn, Science **351**, 576 (2016).
[4] X. Gong, M. Kargarian, A. Stern, D. Yue, H. Zhou, X. Jin, V. M. Galitski, V. M. Yakovenko, and J. Xia, Science Advances **3** (2017).
[5] D. C. Johnston, Advances In Physics **59**, 803 (2010).
[6] J.-H. Chu, J. G. Analytis, K. De Greve, P. L. McMahon, Z. Islam, Y. Yamamoto, and I. R. Fisher, Science **329**, 824 (2010).
[7] Y. Gallais, I. Paul, L. Chauvière, and J. Schmalian, Phys. Rev. Lett. **116**, 017001 (2016).
[8] F. Kretzschmar, T. Böhm, U. Karahasanovic, B. Muschler, A. Baum, D. Jost, J. Schmalian, S. Caprara, M. Grilli, C. Di Castro, J. G. Analytis, J.-H. Chu, I. R. Fisher, and R. Hackl, Nat Phys **12**, 560 (2016).
[9] A. E. Böhmer, P. Burger, F. Hardy, T. Wolf, P. Schweiss, R. Fromknecht, M. Reinecker, W. Schranz, and C. Meingast, Phys. Rev. Lett. **112**, 047001 (2014).
[10] R. M. Fernandes, L. H. VanBebber, S. Bhattacharya, P. Chandra, V. Keppens, D. Mandrus, M. A. McGuire, B. C. Sales, A. S. Sefat, and J. Schmalian, Phys. Rev. Lett. **105**, 157003 (2010).
[11] M. Yoshizawa, R. Kamiya, R. Onodera, Y. Nakanishi, K. Kihou, H. Eisaki, and C. H. Lee, (2010), arXiv:1008.1479.
[12] M. Yoshizawa, D. Kimura, T. Chiba, S. Simayi, Y. Nakanishi, K. Kihou, C.-H. Lee, A. Iyo, H. Eisaki, M. Nakajima, and S. Uchida, Journal of the Physical Society of Japan **81**, 024604 (2012).
[13] F. Weber, D. Parshall, L. Pintschovius, J.-P. Castellan, M. Merz, T. Wolf, and D. Reznik, (2016), arXiv:1610.00099.
[14] D. Parshall, L. Pintschovius, J. L. Niedziela, J.-P. Castellan, D. Lamago, R. Mittal, T. Wolf, and D. Reznik, Phys. Rev. B **91**, 134426 (2015).
[15] S. Avci, O. Chmaissem, J. Allred, S. Rosenkranz, I. Eremin, A. Chubukov, D. Bugaris, D. Chung, M. Kanatzidis, J.-P. Castellan, J. Schlueter, H. Claus, D. Khalyavin, P. Manuel, A. Daoud-Aladine, and R. Osborn, Nat Commun **5**, (2014).
[16] S. Avci, J. M. Allred, O. Chmaissem, D. Y. Chung, S. Rosenkranz, J. A. Schlueter, H. Claus, A. Daoud-Aladine, D. D. Khalyavin, P. Manuel, A. Llobet, M. R. Suchomel, M. G. Kanatzidis, and R. Osborn, Phys. Rev. B **88**, 094510 (2013).
[17] K. M. Taddei, J. M. Allred, D. E. Bugaris, S. Lapidus, M. J. Krogstad, R. Stadel, H. Claus, D. Y. Chung, M. G. Kanatzidis, S. Rosenkranz, R. Osborn, and O. Chmaissem, Phys. Rev. B **93**, 134510 (2016).
[18] L. Wang, F. Hardy, A. E. Böhmer, T. Wolf, P. Schweiss, and C. Meingast, Phys. Rev. B **93**, 014514 (2016).
[19] F. Waßer, A. Schneidewind, Y. Sidis, S. Wurmehl, S. Aswartham, B. Büchner, and M. Braden, Phys. Rev. B **91**, 060505 (2015).
[20] M. Yi, in preparation.
[21] W. R. Meier, Q.-P. Ding, A. Kreyssig, A. Budâko, Sergey L.and Sapkota, K. Kothapalli, V. Borisov, R. ValentÃ, C. D. Batista, P. P. Orth, R. M. Fernandes, A. I. Goldman, Y. Furukawa, A. E. Böhmer, and P. C. Canfield, npj Quantum Materials **3**, 5 (2018).
[22] R. Yu, M. Yi, B. A. Frandsen, R. J. Birgeneau, and Q. Si, (2017), arXiv:1706.07087.
[23] R. M. Fernandes, S. A. Kivelson, and E. Berg, Phys. Rev. B **93**, 014511 (2016).
[24] M. Hoyer, R. M. Fernandes, A. Levchenko, and J. Schmalian, Phys. Rev. B **93**, 144414 (2016).
[25] J. Wang, G.-Z. Liu, D. V. Efremov, and J. van den Brink, (2016), arXiv:1608.07935.
[26] M. H. Christensen, J. Kang, B. M. Andersen, I. Eremin, and R. M. Fernandes, Phys. Rev. B **92**, 214509 (2015).
[27] R. M. Fernandes and A. J. Millis, Phys. Rev. Lett. **111**, 127001 (2013).
[28] T. A. Maier and D. J. Scalapino, Phys. Rev. B **90**, 174510 (2014).
[29] S. Lederer, Y. Schattner, E. Berg, and S. A. Kivelson, Phys. Rev. Lett. **114**, 097001 (2015).
[30] A. E. Böhmer and C. Meingast, Comptes Rendus Physique **17**, 90 (2015).
[31] A. E. Böhmer, F. Hardy, L. Wang, T. Wolf, P. Schweiss, and C. Meingast, Nat Commun **6**, (2015).
[32] In order to properly measure the shear-modulus in this state using our setup, one would have to be able to change the force at low temperature and to measure the bending response, which is unfortunately currently not possible.
[33] A. E. Böhmer, F. Hardy, L. Wang, T. Wolf, P. Schweiss, and C. Meingast, Nat Commun **6**, 7911 (2015).
[34] M. H. Christensen, B. M. Andersen, and P. Kotetes, (2016), arXiv:1612.07633.
[35] L. Wang, F. Hardy, T. Wolf, P. Adelmann, R. Fromknecht, P. Schweiss, and C. Meingast, physica status solidi (b) (2016).